\documentclass[pra,11pt,graphicx,amssymb,floatfix]{revtex4}
\usepackage{graphicx}
\begin{document}

\title{ ``Counterfactual'' quantum protocols }

\author{L. Vaidman}
\affiliation{ Raymond and Beverly Sackler School of Physics and Astronomy\\
 Tel-Aviv University, Tel-Aviv 69978, Israel}

\begin{abstract} \baselineskip 11pt
The counterfactuality of recently proposed protocols is analyzed. A definition of `counterfactuality' is offered and it is argued that an interaction-free measurement of the presence of an opaque object can be named `counterfactual', while proposed ``counterfactual'' measurements of the absence of such objects are not counterfactual. The quantum key distribution protocols which rely only on measurements of the presence of the object are counterfactual, but quantum direct communication protocols are not. Therefore, the name `counterfactual' is not appropriate for recent ``counterfactual'' protocols which transfer quantum states by quantum direct communication.
      \end{abstract}
\maketitle

\baselineskip 14pt
\section{Introduction}

Teleportation \cite{tele} is arguably the most important discovery in quantum information. Teleportation enables the transfer of a quantum state of a system in one site to a system in a remote site.  Teleportation requires a connection between the sites: an  entanglement channel and a classical channel in which  some (surprisingly small) amount of information has to be transferred. Recent ``counterfactual'' protocol \cite{SalihQT,Li} seems to achieve much more: the transfer of a quantum state from one site to another without any quantum or classical particle moving between them. The protocol requires a quantum channel between the sites, but there is only a very small probability that a quantum particle travels in the channel and when it happens, the run of the protocol is discarded.

It is hard to believe that the  counterfactual transfer of a quantum state is possible. It is not just remote preparation, it is a complete teleportation: we need not know the quantum state to be transferred and if the system was entangled, the entanglement is transferred to the  system at the other site. In Sections II-VI I will explain step by step how it works. Section II describes the counterfactual protocol for finding the presence of an object in a remote site (the transfer of bit 1).  Section III describes the counterfactual protocol for finding the absence of this object (the transfer of bit 0). Section IV describes  how using the quantum Zeno effect we  can achieve an arbitrarily high efficiency of these protocols. Section V describes a combined protocol which enables the transfer of both bits 1 and 0 in the same system. Section VI explains how the protocol for transmission of bits can be upgraded to transmission of qubits, i.e. to transmission of quantum states.
But then, in Section VII, I  analyze one of the first steps, the ``counterfactual'' way of finding the absence of an object, and show that it cannot be correct. In Section VIII I propose a natural criterion of counterfactuality which suggests that indeed, the protocol for finding the absence of an object is not counterfactual. In Section IX I discuss which protocols can be considered counterfactual. In Section X I argue that ``counterfactual direct communication protocol'' is not counterfactual. Section XI concludes my analysis.

In this paper I do not provide many details of calculations. Some of the results presented below appear in more detail in \cite{count}, the methods of which can help to reconstruct new results appearing here. Note, that the approach and some of the ideas presented in this work I applied in the past, criticising other ``counterfactual'' protocols \cite{V07,V14}.

\break

\section{Finding an object in  an ``interaction-free'' measurement}\label{IFM}

Penrose \cite{Penrose} coined the term ``counterfactual'' for describing quantum interaction-free measurements (IFM) \cite{IFM}.
In the IFM, an object is found because it might have absorbed a photon, although actually it did not. In the IFM we build a Mach-Zehnder interferometer (MZI) and tune it such that there is complete destructive interference in one of the ports where we place detector $D$, see Fig.~ 1a. We know that that there is nothing inside the interferometer except for one location $O$ at one arm of the interferometer. In this place there might be an opaque object. Now we send a single photon through the interferometer. If detector $D$ clicks, we know that the object is present, see Fig.~1b.

We deduce that the photon was not near the object, because if it were there, it could not reach $D$. I will continue to use this argument in Sections III-VI, but in Section VII I will explain  that this naive classical argument leads to wrong conclusions. In Section IX I will provide another argument why the experiment presented in Fig.~1 {\it is} an interaction-free measurement. However, I see no such alternatives for the cases of Sections III-VI.

\begin{figure} [b]
\begin{center}
 \includegraphics[width=13.2cm]{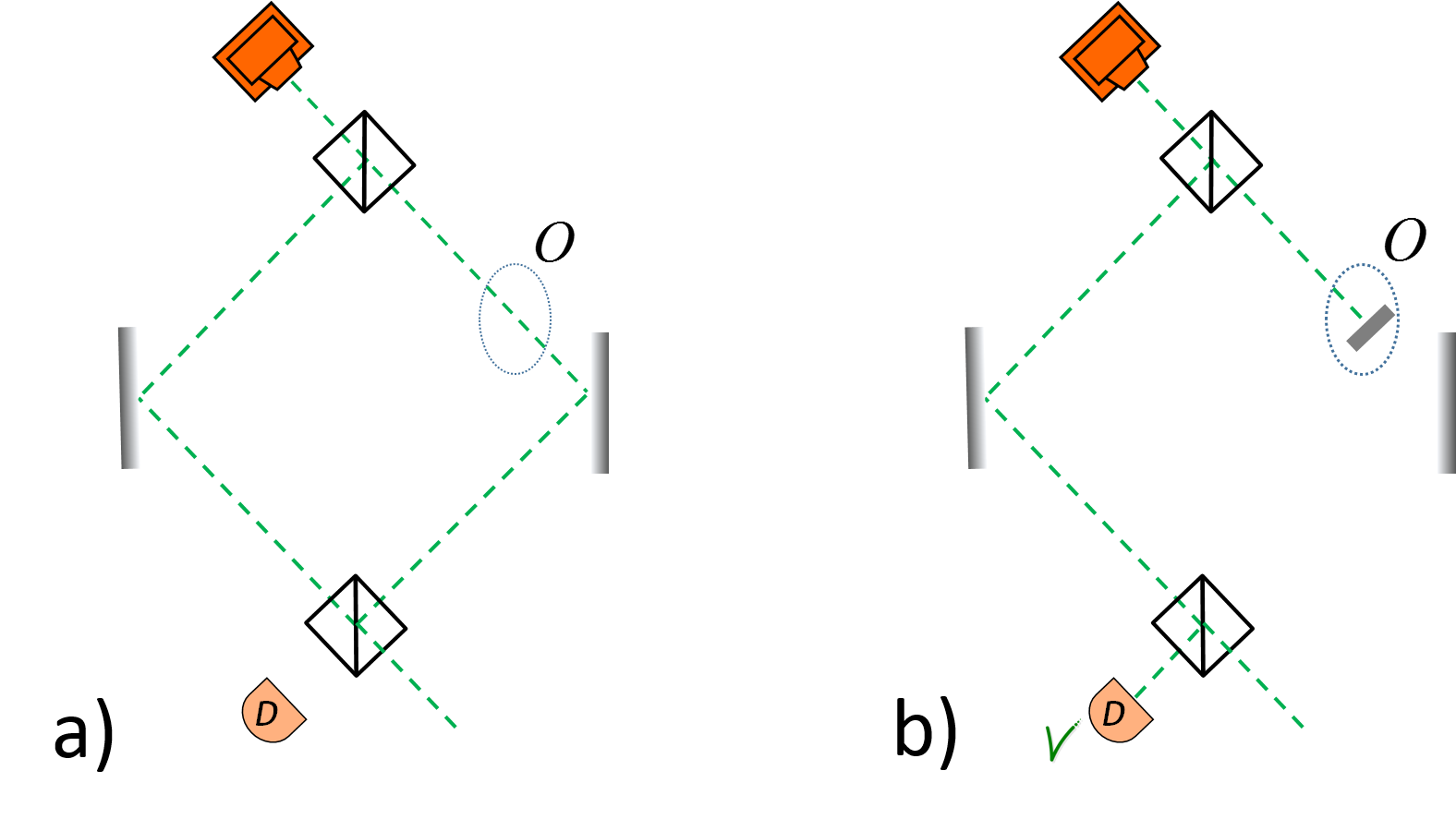}\end{center}
\caption{ \baselineskip 11pt a) The interferometer is tuned in such a way that  detector $D$ never clicks if the paths are free. b)~When detector clicks we  know that the object is in $O$ while photon could not be in $O$. }
\label{fig:1}
\end{figure}

\break

\section{Finding the absence of an object in ``interaction-free'' measurement}\label{IFMabs}

Next step is finding that a particular place is empty. To this end we build and tune a nested MZI, see Fig.~2. The interferometer is tuned (reflectivity of mirrors, distances, etc.) such that two properties are fulfilled: First, the inner interferometer is tuned such that there is a destructive interference toward the beam splitter of the large interferometer, see Fig.~2a. Second, when in the location $O$ there is an opaque object which blocks the arm of the inner interferometer, there is a destructive interference toward the port of the external interferometer with detector $D$, see Fig.~2b. Again, we verify that the interferometer is empty everywhere except for location $O$ in which an opaque object might be present. A single photon is sent through the interferometer. If detector $D$ clicks, we know that the object is not present in location $O$. We deduce that the photon was not present in $O$, because if it were entering the inner interferometer, it could not reach $D$.
\begin{figure}[b]
\begin{center}
 \includegraphics[width=13.2cm]{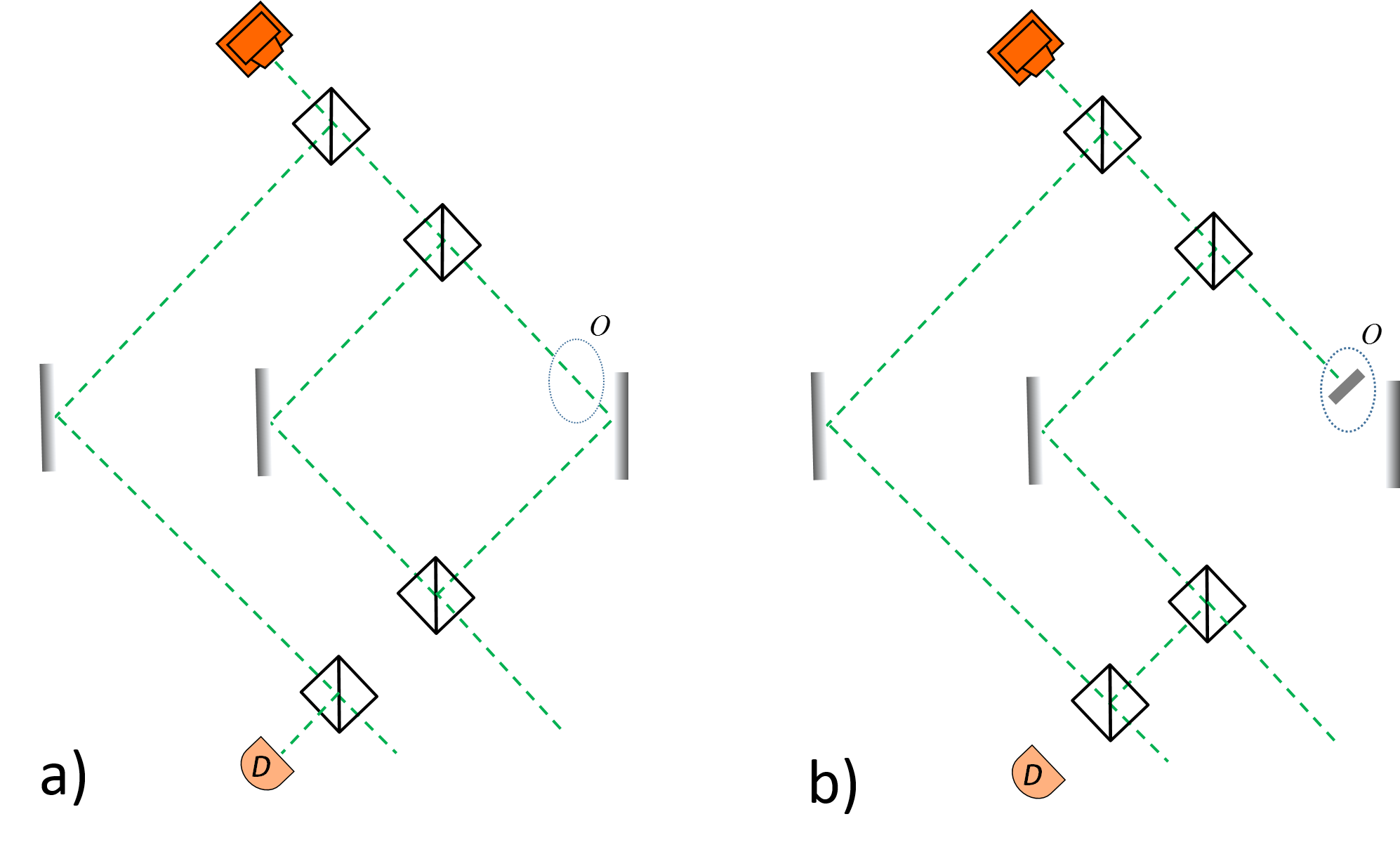}\end{center}
\caption{\baselineskip 11pt a) The inner interferometer is tuned in such a way that when detector clicks, the  photon  could not have been in $O$.  b) The interferometer is tuned in such a way that  detector $D$ never clicks if there is an opaque object in $O$. }
\label{fig:2}
\end{figure}

\break
\section{Adding the quantum Zeno effect}\label{Zeno}

The methods for finding an object and finding that the place is empty described above are robust when we get the click in detector $D$: we get the information with certainty. However, a more probable outcome of these experiments, even if we test for the right thing, i.e. for the presence of the object when it is there, or for the absence of the object when it is not there, is that detector $D$ does not click, which corresponds to a failure of the measurement.
However, these methods can be modified in such a way that the probability of a failure, in case of the test for the correct property, will be arbitrarily small.

 In order to find  an opaque object all we need is a system of two coupled optical cavities: two parallel mirrors and a highly reflecting beam splitter between them \cite{ZenoIFM}. The system is tuned such that a wave packet of a photon  placed in one cavity, after a large number $N$ bounces, ends up being in the other.  However, if in the second cavity we place an opaque object, it stays in the first cavity with probability $\cos^{2N}{\frac{\pi}{2N}} \simeq 1-\frac{\pi^2}{4N}$. The remaining small probability corresponds to the absorbtion  of the photon by the object. Finding the photon in the first cavity tells us with certainty that the object is present.

  In practice, this is how the experiment can be done, but it is easier to discuss the same experiment from a moving frame. Now, the location $O$ transforms into many locations and the object, say a long opaque plate, should be present in all of them, see Fig.~3. We can consider  a stationary chain of $N$ MZIs tuned in such a way that if the interferometer is empty, the photon ends up in the right port, see Fig.~3a, but if the object is present, it ends with high probability in the left port, see Fig.~3b, and is detected by detector $D$.

\begin{figure}[b]
\begin{center}
 \includegraphics[height=22.5cm]{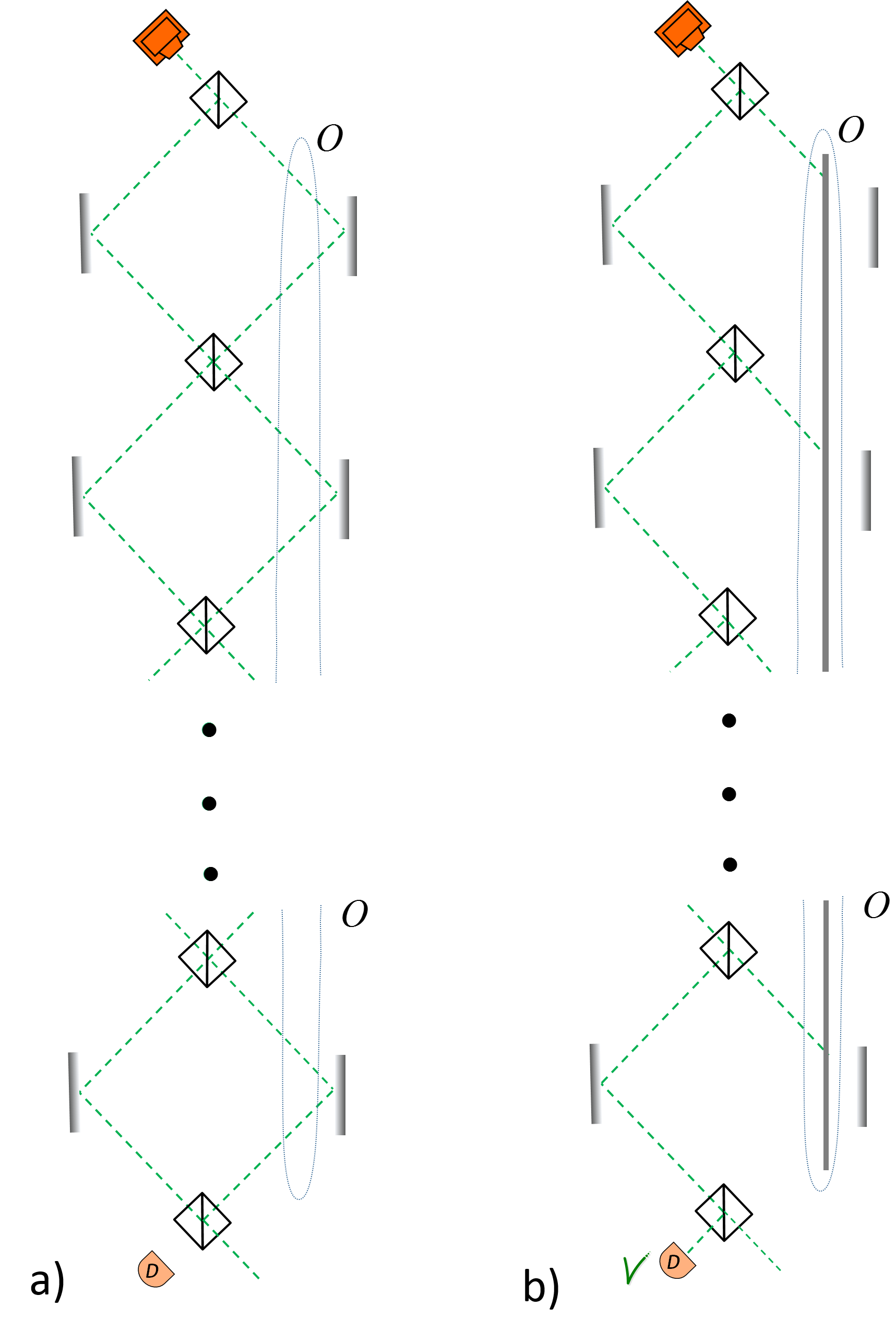}\end{center}
\caption{ \baselineskip 11pt a) The interferometer is tuned in such a way that  detector $D$ never clicks if the paths are free. b)~When detector clicks, we  know that the object is in $O$, while photon could not have been in $O$. Note, that if the object is in $O$, the detector clicks with a very high probability. }
\label{fig:3}
\end{figure}
 A chain of nested interferometers  of the type presented in Fig.~2 with highly reflective central beam splitters can be tuned such that the detector {\it cannot} click if the object is present, but will click with probability close  to 1 if the  interferometer is empty, see Fig.~4. The click of the detector tells us with certainty that the plate is not present in $O$.

\begin{figure}
\begin{center}
 \includegraphics[height=22.5cm]{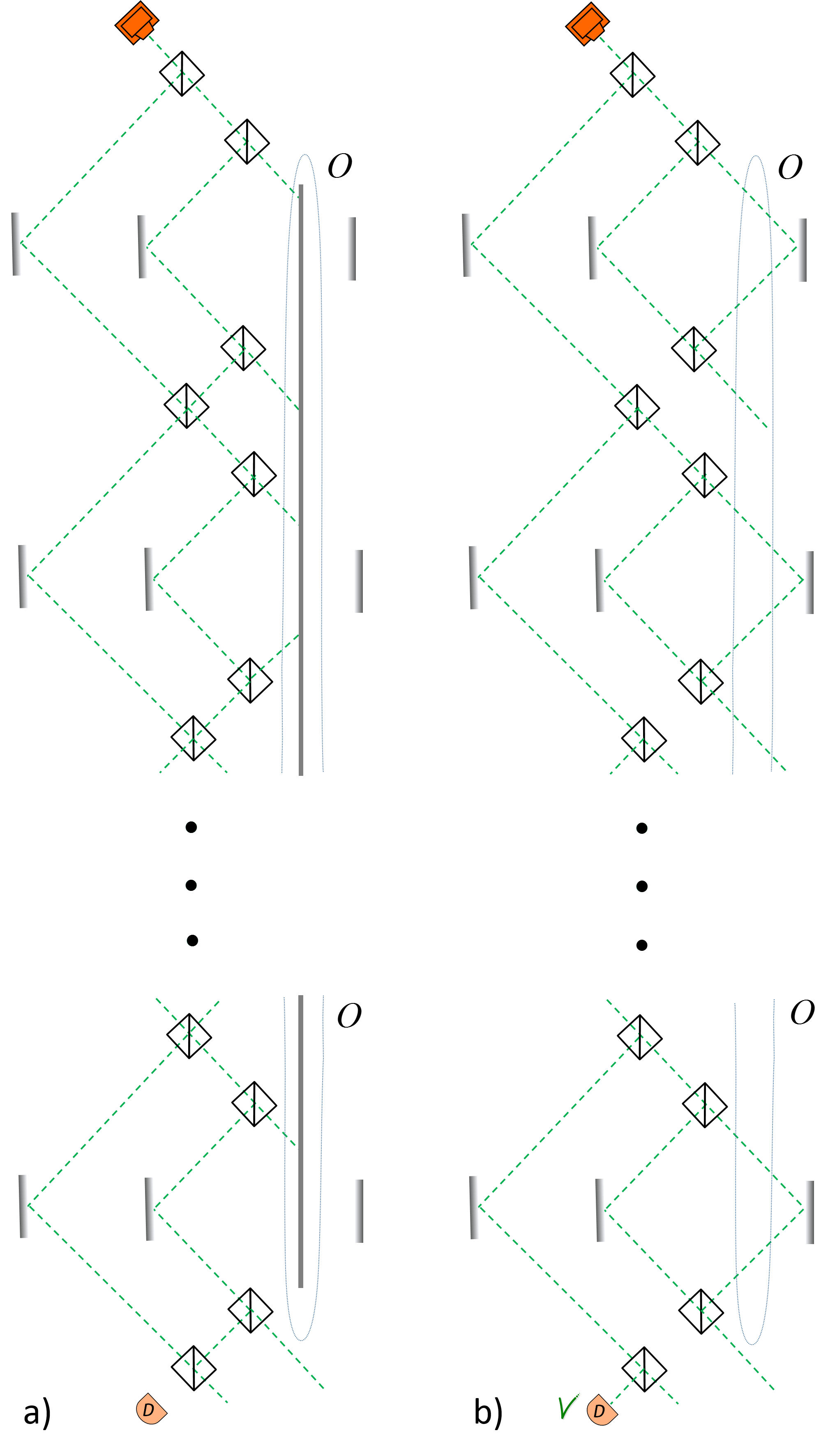}\end{center}
\caption{ \baselineskip 11pt a) The interferometer is tuned in such a way that  detector $D$ never clicks if there is an opaque object in $O$. b) When detector clicks, we  know that the object is not present in $O$ and  photon also could not have been in $O$. Note, that if the place $O$ is empty, the detector clicks with a very high probability. }
\label{fig:4}
\end{figure}

\break

\section{Finding the presence and the absence of an object in ``interaction-free'' measurement}\label{IFMboth}

In the previous section we described a reliable and highly efficient ``interaction-free'' measurement of the presence of an object and another one for the absence of the object. In Fig.~5 a single interferometer is presented which achieves both tasks in the ``interaction-free'' way \cite{Kwiat,Salih}. A chain of large interferometers in each unit of which there is a chain of small interferometers. The inner chains are tuned in such a way that if they are empty, the photon leaves the chain in the right port and does not continue toward  large interferometers. Thus, the small chain acts as an absorber in the right arm of each large interferometer. Therefore, due to Zeno effect, the photon remains on the left side of  large interferometers and ends up in detector $D_1$, Fig.~5a. If, however, all small  interferometers are blocked, the Zeno effect in small interferometers prevents losing the photon and, due to proper tuning of the large chain, the photon ends up in detector $D_2$. In both cases we deduce that the photon, if detected by $D_1$ or $D_2$ (which happens with probability close to 1), could not be in location $O$ because if it were there, it could not reach the detectors.

This is a counterfactual direct communication between Bob who can place (bit 1) or not place (bit 0) the plate and Alice who sends the photon into the interferometer and obtains the information by observing a click in detector $D_2$ (bit 1) or $D_1$ (bit 0). This is  an ``interaction-free'' transfer of a  classical bit. No particles moved between Bob and Alice in this process. This process was named as ``counterfactual computation'' \cite{Kwiat} and ``counterfactual communication'' \cite{Salih}.

\begin{figure}
\begin{center}
 \includegraphics[width=11.6cm]{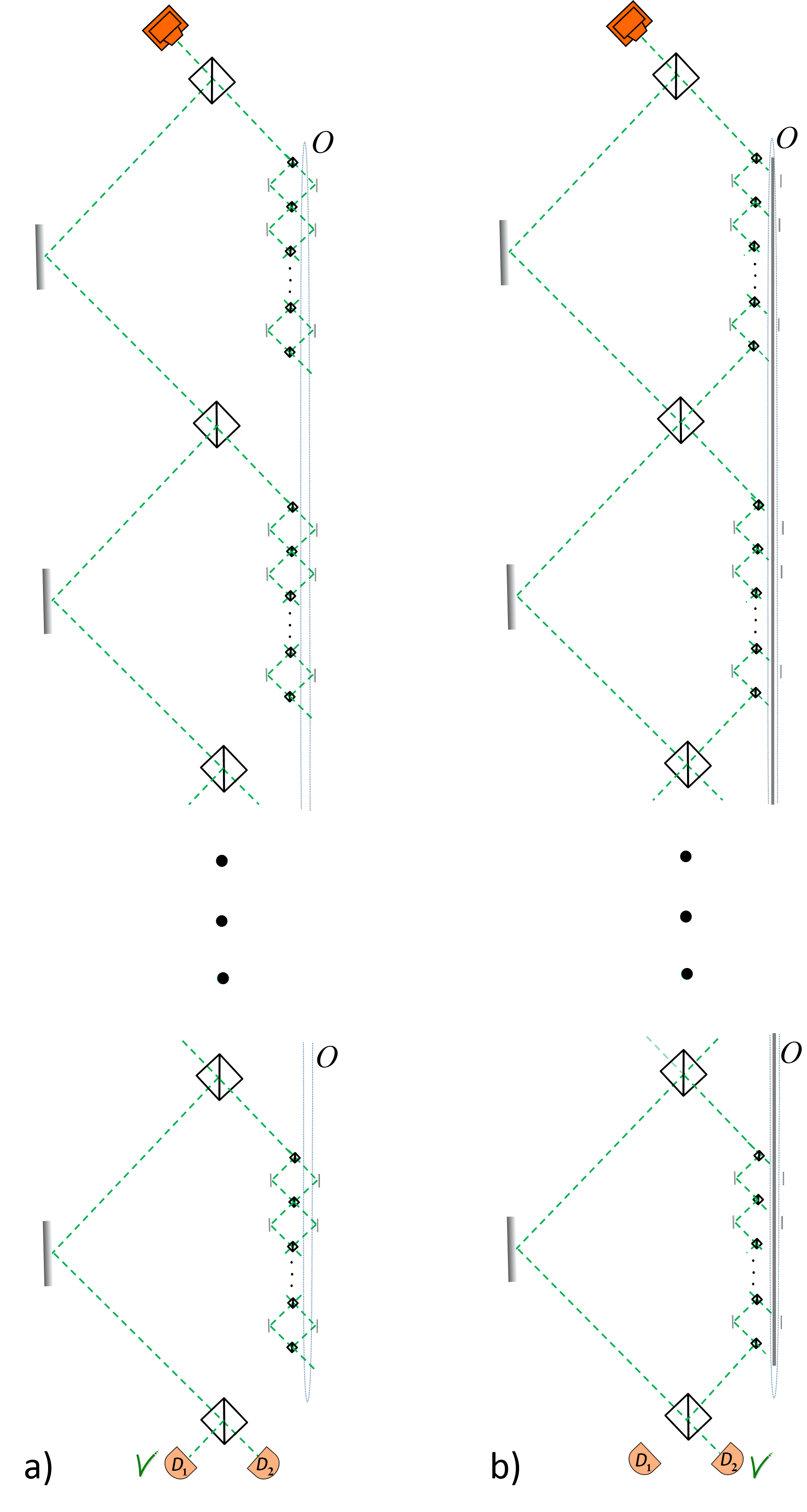}\end{center}
\caption{\baselineskip 11pt a) The interferometers in small chains are tuned in such a way that the photon entering any such chain does not return to the large interferometer, so the chain acts as an absorber. Thus, when the interferometer is empty, $D_1$ clicks with high probability and in this case the photon could not have been in $O$.  b) When the object is present in $O$,   every small chain becomes a highly reflective mirror. The large interferometers are tuned in such a way that the photon cannot leave toward $D_1$. The click of $D_2$, which happens with high probability,
tells us that the object is in $O$ while photon could not have been  in $O$.}
\label{fig:5}
\end{figure}

\break

\section{``Interaction-free'' transfer of a quantum state}\label{IFMboth}

Recently, based  on the above method for ``interaction-free'' transfer of a bit, an ``interaction-free'' method for transmitting a qubit was proposed  \cite{SalihQT,Li}. If we consider the presence and the absence of the plate as two quantum states of a system on Bob's side, then the procedure provides the following transformation:
\begin{eqnarray}\label{U1}
  |R\rangle_A |1\rangle_B &\rightarrow&  |1\rangle_A |1\rangle_B, \\
|R\rangle_A |0\rangle_B &\rightarrow&  |0\rangle_A |0\rangle_B .\label{U2}
\end{eqnarray}
where $|R\rangle_A$, the ``ready'' state,  is the photon wave packet entering the interferometer on Alice's side, $|0\rangle_A$ and $|1\rangle_A$ are states exiting the interferometer toward detectors $D_1$ and $D_2$ respectively, $|0\rangle_B$ and $|1\rangle_B$ are states of the opaque object on Bob's side.

The linearity of quantum mechanics tells us that if Bob's plate is prepared in a superposition:
\begin{equation}\label{sup}
 |\psi\rangle_B=\alpha |1\rangle_B + \beta |0\rangle_B ,
\end{equation}
then the procedure  creates   entanglement between Bob's and Alice's systems \cite{Guo}:
\begin{equation}\label{Uent}
 |R\rangle_A|\psi\rangle_B \rightarrow   \alpha |1\rangle_A |1\rangle_B  + \beta |0\rangle_A|0\rangle_B .
\end{equation}

The time symmetry of quantum processes tells us that process (\ref{Uent}) can be reversed. Bob's and Alice's systems will be disentangled and the quantum state of Bob's system will be restored.
Equation (\ref{Uent}) shows that the entangled state obtained after the process is symmetric between Alice and Bob, so if in this  reversal operation the roles of Alice and Bob are switched, the state  $|\psi\rangle$ will end up in Alice's hands:
\begin{equation}\label{U-1}
 \alpha |1\rangle_A |1\rangle_B  + \beta |0\rangle_A|0\rangle_B \rightarrow   |\psi\rangle_A |R\rangle_B .
\end{equation}
This is a gedanken argument because we consider Alice's photon and Bob's plate on the same footing and assume the existence of technology which can make an interference experiment with Bob's plate. The state $|R\rangle_B $ is a quantum state of Bob's plate exiting  Bob's interferometer toward its input port in the reversal operation. A conceptually equivalent, but   more realistic proposal is described in \cite{Li}.

Given that the operations (\ref{U1}) and (\ref{U2}) are interaction-free, the process for creating entanglement (\ref{Uent}) is also interaction-free, as well as  the reverse operation (\ref{U-1}). Thus, the two-step operation provides the ``interaction-free'' transfer of a quantum state from  Bob to Alice without particles traveling between them.

I explained how to transfer a quantum state from Bob to Alice. It can be an unknown quantum state and it can be an entangled quantum state, the transfer works as in a teleportation protocol, but with less resources: no need for prior entanglement and no need to send classical information. Note, however, that the process takes more time than quantum teleportation, so  relativistic causality is not broken. Still, it is amazing, revolutionary, and unbelievable!  Indeed, I do not believe that it is true \cite{Mycom}. In  the next section I  present an argument why it cannot be true.

\break

\section{Why ``interaction-free'' measurement of the absence of the object cannot be interaction free}\label{noIFMabs}

One of the crucial steps for counterfactual transfer of a quantum state was ``interaction-free'' measurement which told us that a particular place is empty, see section \ref{IFMabs}. The method was presented in Fig.~2. Consider a successful experiment with a click of detector $D$, see Fig.~6a. We deduce that there is no  opaque object in $A$  and  that the photon was not there either (i.e. it was an ``interaction-free'' measurement) because if it were there, it could not reach the detector. By the same logic we can argue that it was not in $B$. We also can argue that since it entered and left the interferometer, it was somewhere in the arms $A$, $B$ and $C$. Therefore, the photon was in arm $C$. This is case a).

In a properly tuned similar interferometer, see Fig.~6b.,  we can claim that the photon  was in arm $A$ and was not in arms $B$ and $C$. This is case b).

These claims rely on classical reasoning. If we believe that the true description of Nature is quantum mechanical, we must reach the same conclusion if we consider the quantum description of the photon. I will show, however, that this is not the case.
 Indeed,   both in case a) and in case b), the quantum state of the photon transforms from the entrance to the middle part of the interferometer in the same way (see \cite{Danan}):
\begin{equation}\label{ABC}
|\Psi_{in}\rangle \rightarrow \frac{1}{\sqrt 3}(|A\rangle+i|B\rangle+|C\rangle).
\end{equation}

\begin{figure}[b]
\begin{center}
 \includegraphics[width=12.2cm]{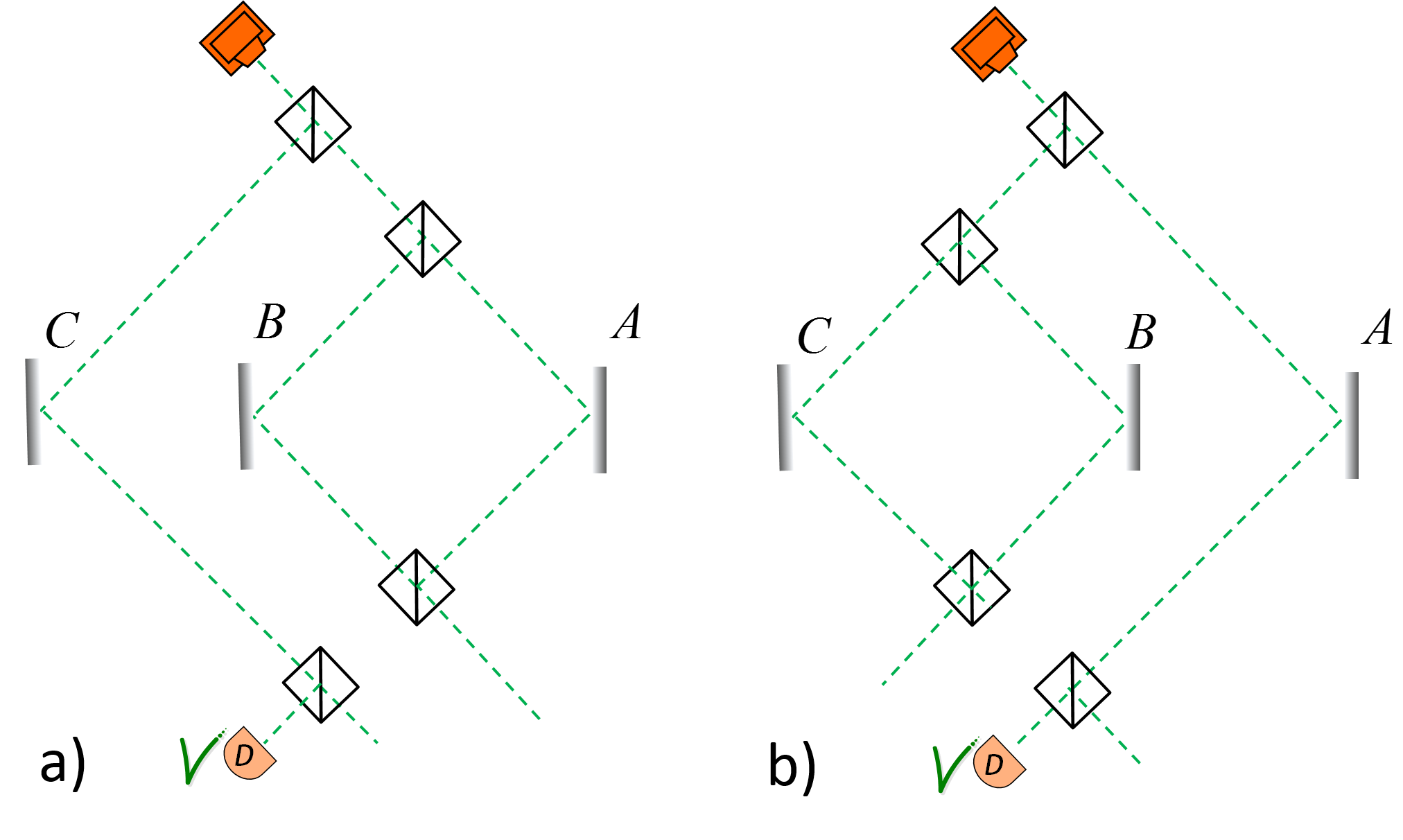}\end{center}
\caption{ \baselineskip 11pt a) The interferometer is tuned as described in Fig.~2. The case of the click of  detector $D$ is considered. A naive classical argument tells us that the photon was solely in  path $C$. b) Similar configuration with nested interferometer  in the other arm of the large interferometer.  In this case we deduce that the photon took path $A$.}
\label{fig:6}
\end{figure}

\break

Another relevant information is how each one of the states in the middle of the interferometer evolves toward the output. These evolutions  have the same form for case a) and case b):
\begin{eqnarray}\label{ABC2}
 \nonumber
 |A\rangle &\rightarrow& \frac{1}{\sqrt 3} |D\rangle+{\sqrt \frac{2}{3}}|\bot_A\rangle ,\\
 |B\rangle &\rightarrow& i\frac{1}{\sqrt 3} |D\rangle+{\sqrt \frac{2}{3}}|\bot_B\rangle ,
 \nonumber \\
 |C\rangle &\rightarrow& \frac{1}{\sqrt 3} |D\rangle+{\sqrt \frac{2}{3}}|\bot_C\rangle ,
 \end{eqnarray}
where state $|D\rangle$ signifies a wave packet entering detector $D$, and $|\bot_A\rangle$  is a normalized remaining part of the wave function evolving from state $|A\rangle$, etc.
The states $|\bot_A\rangle$, $|\bot_B\rangle$, and  $|  \bot_C\rangle$ are different for the cases a) and b), but the scalar products $\langle \bot_A|\bot_B\rangle$, $\langle \bot_A|\bot_C\rangle$, etc. are all the same, and this is what I  expect to be relevant.

 There is no difference between quantum descriptions of cases a) and b), both are  symmetric between $A$ and $C$.  However, the naive classical analysis tells us that the photon was in $C$ in case a) and in $A$ in case b).

 The quantum description of the photon in the middle of the interferometer (in both cases) is exactly the same as in the so called ``3-box paradox'' \cite{AV91}. Given the pre- and post-selection, the photon is found with certainty in $C$ if it is searched there, and also it is found with certainty in $A$ if it is searched there instead. There is a complete symmetry between $C$ and $A$, so there is no way to argue that the photon is in $C$, but not in $A$.

\section{Definition of the ``interaction-free'' process}\label{IFMdef}

I assume that all interactions are local. So if a particle could not have been in the vicinity of an object, I claim that the particle had no interaction with this object. But ``being near the object'' is a classical concept. The  quantum particle is described by a quantum  wave function, part of which {\it was} near the object. The pre-selected photon, the photon which  entered the interferometer is present in all three arms. It is the additional information about the post-selection (detection by detector $D$) which led us to tell that  the particle took path $C$. We have seen that the naive argument, according to which the particle was in the path through which it could reach the detector, runs into a contradiction. My proposal, instead, is to say that the pre- and post-selected particle was where it left a trace \cite{past}.

Since ``trace'' is a consequence of interaction, it by definition   helps us to know if there was an interaction between the particle and the object: If the particle left no trace near the object it is an interaction-free process. The interaction-free measurement that a particular place is empty then requires that the particle will leave no trace in this place.

 A very useful framework for finding places where a pre- and post-selected particle leaves a trace is the two-state vector formalism (TSVF) \cite{AV90}. In this framework the pre- and post-selected quantum system is described by a pair of quantum states, the usual one defined by pre-selection and the backward evolving state defined by the post-selection. The TSVF provides a simple criterion for finding the trace of a pre- and post-selected particle. The trace is present only in the overlap of the forward and the backward evolving wave functions.

\begin{figure}[t]
\begin{center}
 \includegraphics[width=12.2cm]{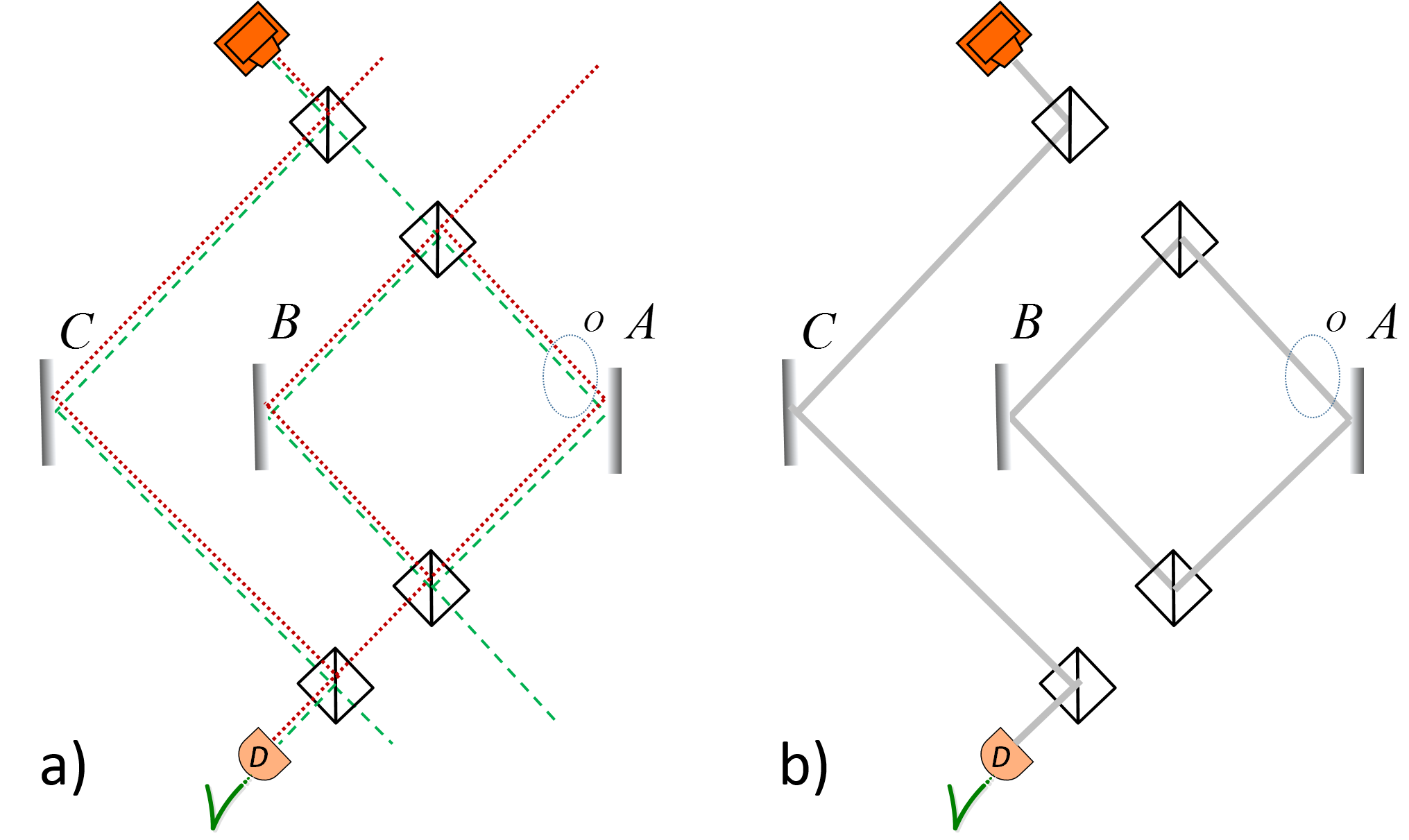}\end{center}
\caption{\baselineskip 11pt a) The forward evolving state of the photon is represented by a dashed line, while the backward evolving state created by the click of the detector is represented by a dotted line. The overlap is on path $C$ and inside the inner interferometer. b) The weak trace in the environment left by pre- and post-selected photon.  }
\label{fig:7}
\end{figure}

Consider the ``interaction-free'' experiment for finding that there is nothing in place $O$  when it is successful, i.e.   detector $D$ clicks, see Fig.~6a. The forward and backward evolving states are shown in Fig.~7a. The TSVF tells us that  there is a trace in $O$, see Fig.~7b. According to my definition, the photon was there, and therefore this process is not interaction-free. Note, that in the middle of the interferometer the overlap is in all arms, the picture is symmetric as it is suggested by the quantum mechanical description.

My other reason to claim that the photon was both in $C$ and in $A$ (and also in $B$) is the experiment we performed in Tel Aviv \cite{Danan} in which we ``asked'' the photons in the interferometer: where they have been? We introduced small disturbances which have different characteristics in five places inside the interferometer. The photons detected in $D$ showed equally three of them, telling us that the photons were in $A$, $B$ and $C$. This is essentially a weak measurement of the trace left by the photons, but since it is very difficult to measure it using external measuring devices, the pointer variable was the transversal degree of freedom of the photons themselves.

\section{Truly ``interaction-free'' protocols}\label{IFMdef}

If we apply the ``trace'' definition of the interaction-free character of a process to the original ``interaction-free measurement'', we will see that it remains interaction free. Indeed, in finding the object in the interaction-free method described in Fig.~1, there is no overlap of forward and backward evolving waves near the object and there is no trace of the particle near the object, see  Fig.~8.  Thus, the key distribution protocol \cite{Noh}, which is based solely on interaction-free finding of an object, can be named counterfactual.

\begin{figure} [h]
\begin{center}
 \includegraphics[width=12.2cm]{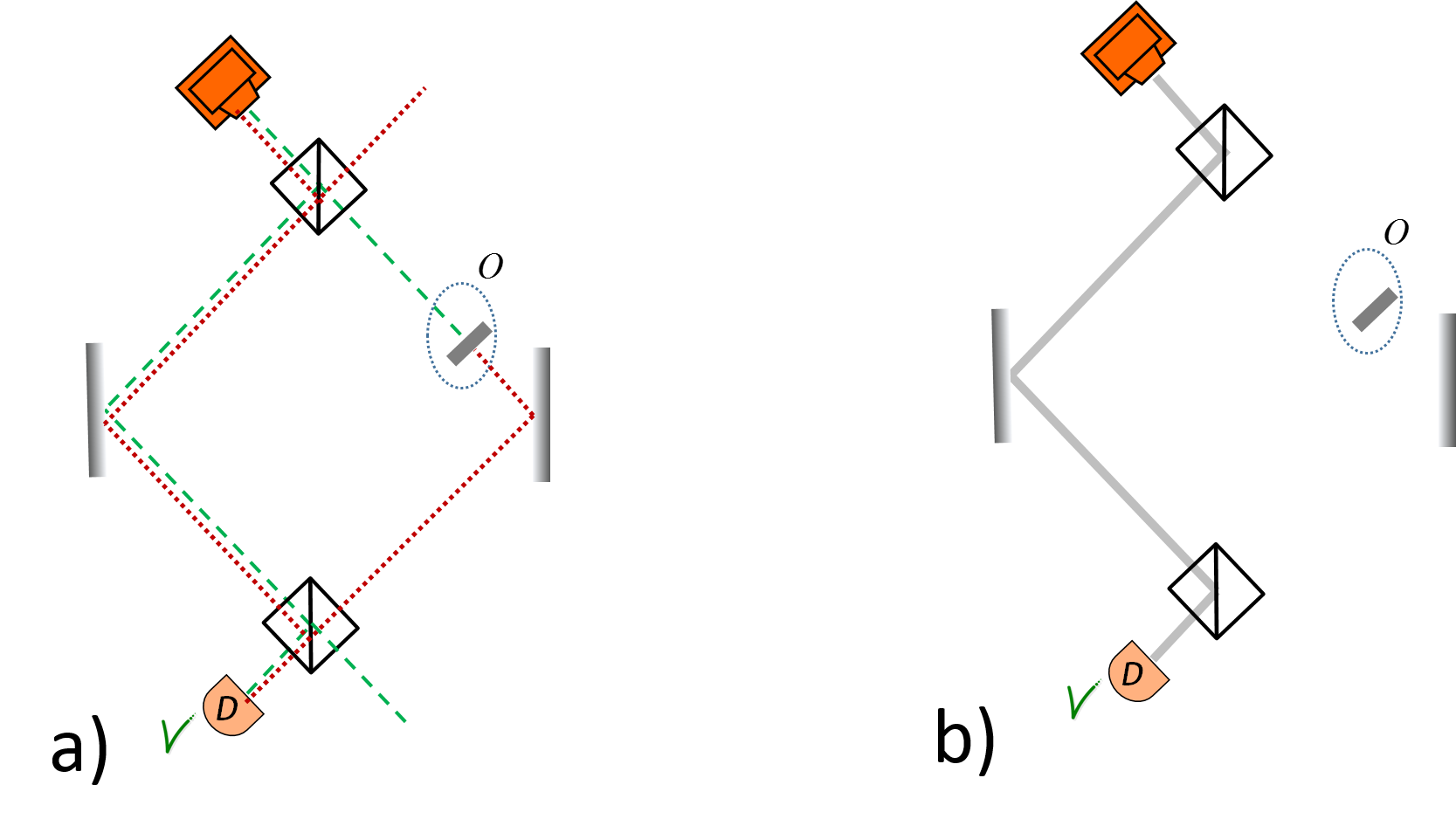}\end{center}
\caption{ \baselineskip 11pt Interaction-free measurement. a) The forward evolving state of the photon is represented by a dashed line, while the backward evolving state, created by the click of the detector, is represented by a dotted line. The overlap is only in the left arm  of the interferometer. b)~The weak trace in the environment left by the pre- and post-selected photon.  }
\label{fig:8}
\end{figure}

Formally, the latest experiment on counterfactual computation \cite{K-Du} also can be considered to be interaction-free.
    But this is because it demonstrates the ``generalized counterfactual computation protocol'' \cite{Mi-Jo} which is different from the counterfactual computation as it  was originally defined. In the ``generalized counterfactual protocol'', the property is not that the particle is not present near the object, but that it is not absorbed by the object. The task is to find the only empty channel out of $N$ channels without absorbing the photon in any of the channels.  It is known that $N-1$ channels are blocked by opaque objects and one is free. I described above how an opaque object can be found without absorbtion. This allows to find all blocked channels without absorbing the particles by these objects.
     However, it seems to me misleading to name it ``counterfactual'' because photons are present in the empty channel which we want to discover.

   \section{ ``Counterfactual'' communication}\label{IFMdef}

   The ``counterfactual'' transfer of a quantum state and the ``counterfactual'' communication is not interaction-free according to my definition.  The particle leaves a trace when the bit is 0, in the place where the object is absent, so I do not want to name this process counterfactual. The main argument for naming these protocols ``counterfactual'' was that the photon was not present in the transmission channel. Fig.~9 shows the location of the weak trace when the object is present and Fig.~10 shows the location of the weak trace when the object is absent. There is no way to define Alice's and Bob's territories such the there will be a place between them without trace for both cases. So, there is no way to claim that the photon was not in the transmission channel in these protocols.

\begin{figure}
\begin{center}
 \includegraphics[width=11.6cm]{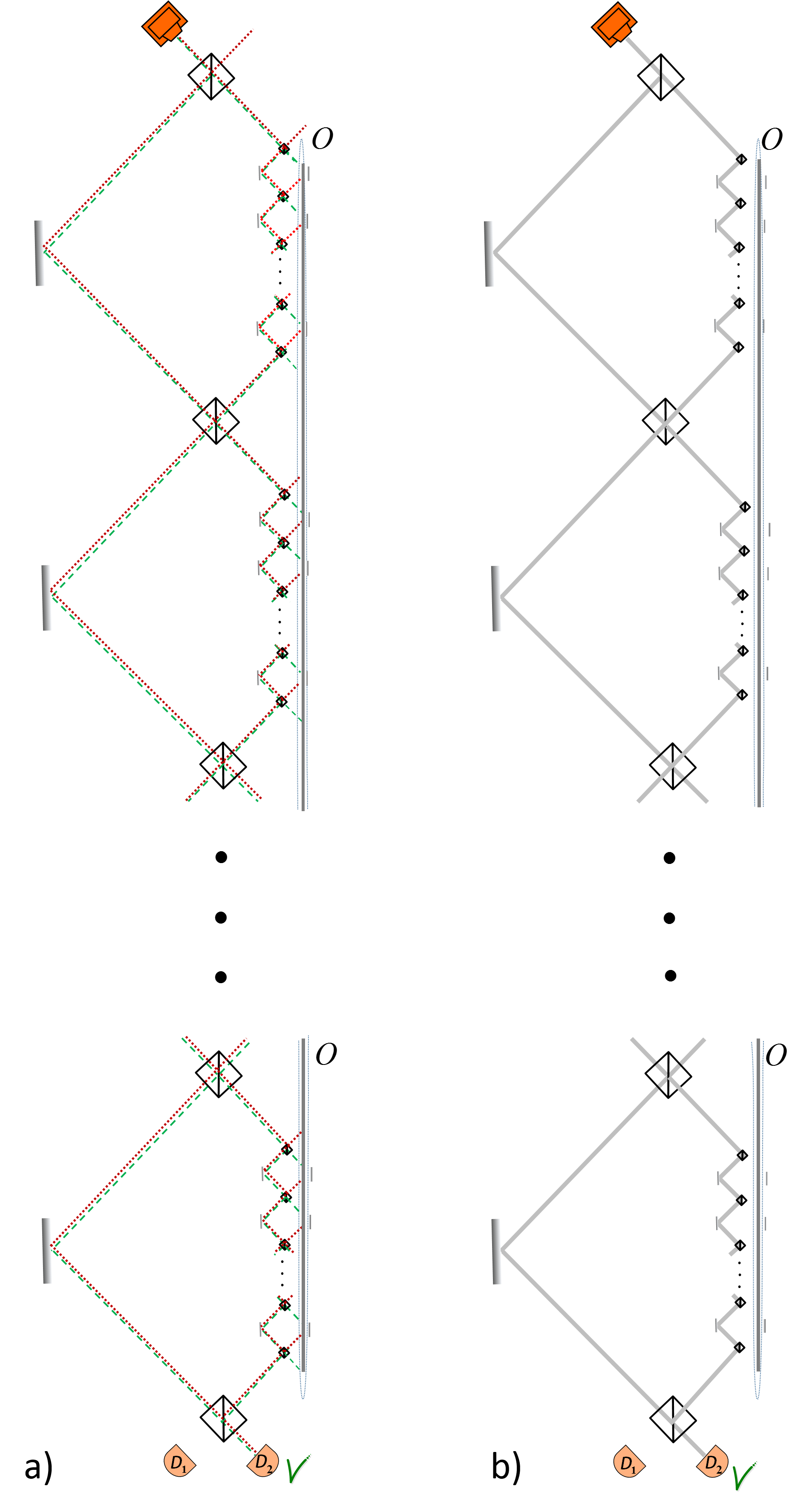}\end{center}
\caption{\baselineskip 11pt ``Counterfactual'' communication of bit 1. a) The forward evolving state of the photon is represented by a dashed line, while the backward evolving state, created by the click of the detector, is represented by a dotted line. b) The weak trace in the environment left by the pre- and post-selected photon. It is not present in Bob's territory which can be defined just as location $O$. }
\label{fig:9}
\end{figure}

\begin{figure}
\begin{center}
 \includegraphics[width=11.6cm]{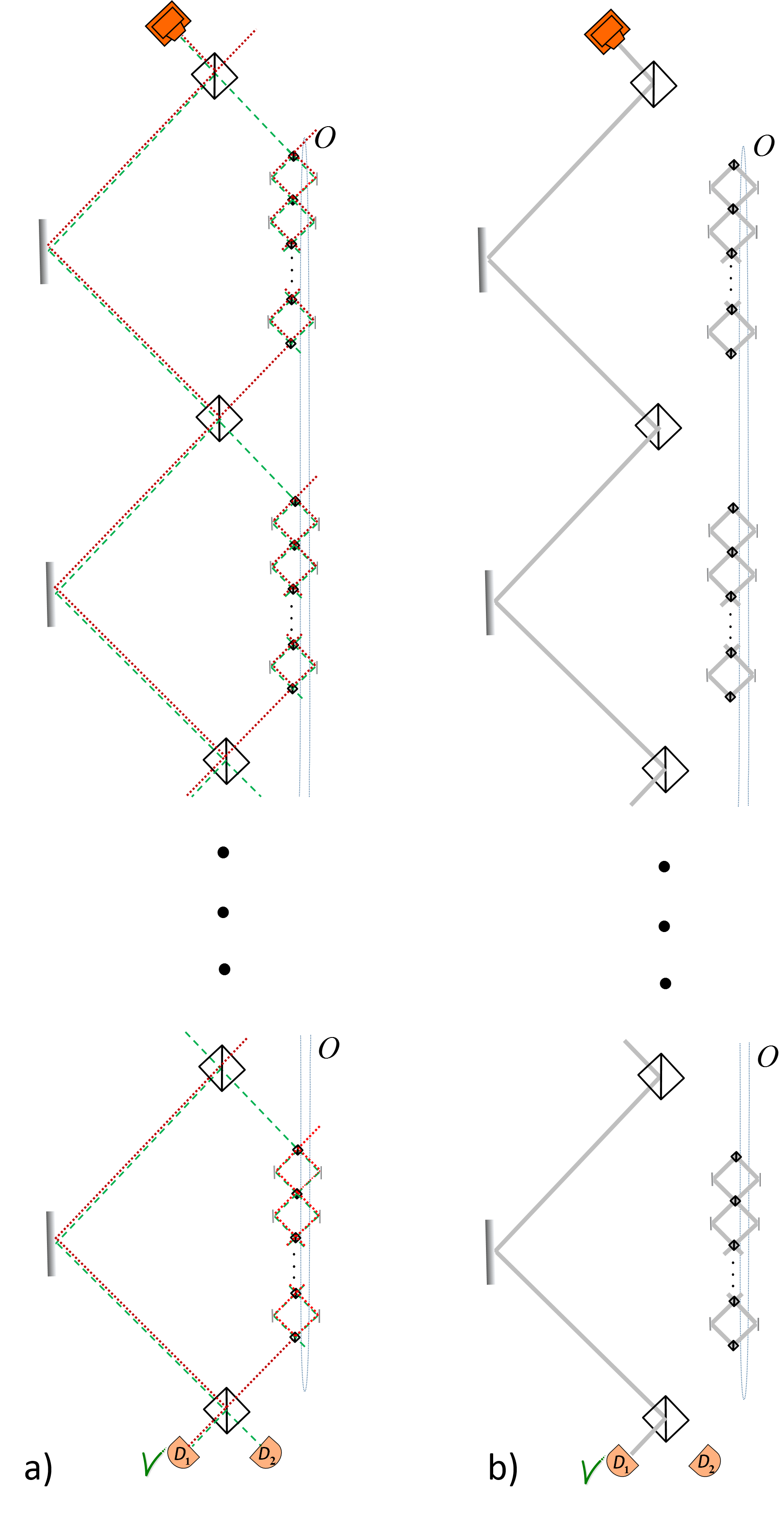}\end{center}
\caption{\baselineskip 10pt ``Counterfactual'' communication of bit 0. a) The forward evolving state of the photon is represented by a dashed line, while the backward evolving state, created by the click of the detector, is represented by a dotted line. b) The weak trace in the environment left by the pre- and post-selected photon. It is present in Bob's side, but there is some area between Alice and Bob without the trace. }
\label{fig:10}
\end{figure}
   If we perform a protocol for direct ``counterfactual'' communication of classical bits, we can argue that the photons never {\it cross} the transmission channel. Fig.~9  and Fig.~10 show that for every bit there is a (different) part of the channel without trace.  Thus, making definition of a protocol as counterfactual when no particles {\it cross} the transmission channel, justifies the name ``counterfactual''. In a somewhat artificial way it also can be named ``interaction-free'', since when the object is present on Bob's side, the photons are not present there, and when the photons are present on Bob's side, the object is not there.

   Even this limited definition of ``counterfactual'' does not allow counterfactual transfer of a quantum state. When Bob's object is in a superposition of being present and absent in location $O$, every photon in the interferometer leaves a trace in all parts of the channel. (Note that the trace has an interesting entanglement feature: if it is identified in one place, it will disappear in the other.)

    \section{Conclusions}\label{IFMdef}

In my view, ``counterfactual'' transfer of a quantum state is far from being on a par with teleportation. The claim that the transfer is achieved without particles in the transmission channel relies on a naive classical argument which is not applicable for quantum systems. Accepting the definition that the particle is present where it leaves a trace makes this protocol not counterfactual.

The  direct ``counterfactual'' communication protocol has more merit for the name counterfactual since even with the trace definition of the presence of the particle we can claim that in the proper operation of the protocol there are no particles crossing the transmission channel.  But they somehow appear on the other side of the channel. This prevents me to accept counterfactuality of this protocol too.

The trace definition leaves fully counterfactual the ``counterfactual'' key distribution protocols. They rely on ``interaction-free'' detecting of an opaque object which is counterfactual according to the trace definition.

I have shown that  counterfactuality of ``counterfactual'' protocols is limited. But even the ``truly  counterfactual'' protocol for interaction-free detection of an opaque object I find counterfactual only subjectively, for us. I believe that there should be a local explanation for everything, so this, and other quantum phenomena lead me to accept the many-worlds interpretation of quantum mechanics according to which the interaction-free measurement  is interaction free only in our world. On the level of the physical universe which incorporates all worlds together, it is not: there is an interaction in parallel worlds \cite{MWI}.

I thank Kelvin McQueen for helpful discussions.
This work has been supported in part by the Israel Science Foundation  Grant No. 1311/14  and the German-Israeli Foundation  Grant No. I-1275-303.14.

\end{document}